\documentclass[twocolumn]{elsart}
\usepackage{array,dcolumn}
\usepackage{graphicx} 
\usepackage{epsfig}
\newcommand{\micron}{$\mathrm{\mu m}$\ }
\newcommand{\tsup}[1]{\textsuperscript{#1}}
\newcommand{\nequ}{$\mathrm{n_{eq}}$}

\begin{document}
\begin{frontmatter}
\title{A Measurement of Lorentz Angle of Radiation-Hard Pixel Sensors}
\author[Milano]{Mario Aleppo \thanksref{apc}}
\address[Milano]{Dipartimento di Fisica,
Universit\`{a} di Milano and
INFN, Sezione di Milano,
Via Celoria 16, I-20133 Milano, Italy}
\thanks[apc]{On Behalf of the ATLAS Pixel Collaboration\cite{TDR}}
\begin{abstract}
Silicon pixel detectors developed to meet LHC requirements were tested
in a beam at CERN in the framework of the ATLAS collaboration.
The experimental behaviour of irradiated and not-irradiated sensors 
in a magnetic field is discussed. 
The measurement of the Lorentz angle for these sensors at different 
operating conditions is presented.
A simple model of the charge drift in silicon before and after irradiation
is presented. The good agreement between the model predictions and 
the experimental results is shown.
\end{abstract}
\end{frontmatter}
\date{29 August 2000}
\section{Introduction}
In the presence of an electric field ${\boldmath E}$ and a magnetic field 
${\boldmath B}$ mutually orthogonal the charge carriers 
move along a direction that forms an angle $\gamma$ ( Lorentz angle ) 
with the electric field. 
This angle affects the area of collection of the charge carriers.

Resolution and efficiency of the detector depend on 
the track incidence angle and on the charge drift angle: the determination 
of this angle is therefore very important to define the mechanical
design and optimize detector performance.    

The silicon pixel detector of the ATLAS experiment will be exposed at 
intense fluxes of radiation during its lifetime; as a result the 
properties and the conditions of operation of the sensors will change. 
Several single chip assemblies 
were characterised extensively in test beam experiments 
performed at the CERN SPS accelerator with a pion beam of 
180 GeV/{\em c} momentum~\cite{ragusa}. 

A beam telescope consisting of 4 pairs of silicon microstrips detectors
(each pair consisting of two planes of detectors with orthogonal strips)
was used to measure the transverse position of the
incident beam particles. 
The pixel
assemblies and the silicon strip telescope were positioned inside
a magnet for the measurement of the Lorentz angle.
 
\subsection{ATLAS Pixel Sensors}
A detailed description of the ATLAS pixel sensors can be found 
in~\cite{TDR,sensors}, so only a brief description is given here.
The sensor is a matrix of 18~$\times$~160 {\it n$^+$}
implants on a high resistivity {\it n}-bulk substrate.
This choice allows for operation in partially depleted 
mode after bulk inversion induced by radiation damage.
The isolation between {\it n$^+$} pixels is obtained using the newly developed 
{\it p-spray} technique~\cite{p-spray}. 
The dimensions of pixel cell are 50~\micron~$\times$~400~\micron.
The pulse height measurement is performed by measuring the
time the pulse from the amplifier remains above the threshold (Time
Over Threshold). 
Typical thresholds were around 3000 electrons.
Some detectors were exposed to a fluence comparable to those expected for LHC. 
Irradiations were performed using the 300 MeV/$c$ pion beam at PSI and the 55
MeV/$c$ proton beam at LBNL.
Sensors irradiated with fluences of 0.5$\times$10\tsup{15} and
1$\times$10\tsup{15} \nequ/cm\tsup{2}, hereafter referred respectively
as half-irradiated and full-irradiated sensors, were 
tested in the beam.  
They were cooled at $-9^{0} \mathrm{C}$ during data taking.

\section{Lorentz angle measurement}

The distribution of charge produced by ionizing particles  
while drifting to the read-out pixels can spread over more than one pixel. 
The spread depends on the particle incidence angle and 
is minimum for an angle equal to the Lorentz angle. 

The Lorentz angle was extracted finding the minimum 
(with a parabola fit) of the mean 
cluster size measured as a function of the angle of the incident 
beam particles with respect to the normal to the detector~\cite{Pioneer}.   
For each angle data were taken both without magnetic field and with
a magnetic field of 1.4 Tesla. 
Data taken without magnetic field were used to check 
the consistency of the measurement with an expected value 
of $0^0$. 

\begin{table*}[htb!]
\begin{center}
{\footnotesize
\caption{Lorentz angle measurement results.}
\label{table:2}
\newcommand{\cc}[1]{\multicolumn{1}{c}{#1}}
\renewcommand{\arraystretch}{1.2} 
\begin{tabular}{ | c | c | c | c | c | }
\hline  
Fluence             &  Volts &  Dep. depth [$\mu$m] & $\theta_L[{}^0] $(Parabola) & 
$\theta_L[{}^0]$ (Model) \\
\hline
 0                  &  150    &     $283 \pm 6$   &  $9.0 \pm 0.4 \pm 0.5$ & $  8.6 \pm 0.4 $ \\

$5\times 10^{14}$   &  150    &     $123 \pm 19$  &  $5.9 \pm 1.0 \pm 0.3$  & $ 5.3 \pm 0.5 $ \\

$5\times 10^{14}$   &  600    &     $261 \pm 8$   &  $2.6 \pm 0.2 \pm 0.3$  & $ 3.9 \pm 0.2 $ \\

$10^{15}$ ('98)  &   600    &     $189 \pm 12$     &  $3.1 \pm 0.4 \pm 0.6$  & $ 2.9 \pm 0.2 $ \\

$10^{15}$ ('99)  &   600    &     $217 \pm 13$     &  $2.7 \pm 0.4 \pm 0.4$  & $ 3.2 \pm 0.3 $ \\

\hline
\end{tabular}
}
\end{center}
\end{table*}

\begin{figure}[htb!]
  \begin{center}
    \leavevmode
    \setlength{\unitlength}{1.0mm}
\includegraphics[clip,scale=0.3]{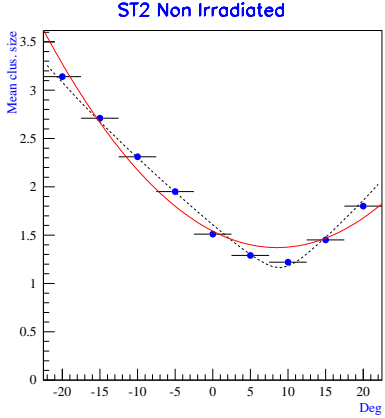}
  \end{center}
\vspace{-.5cm}
\caption{Mean cluster size as a function of the track incidence angle in 
a magnetic field of 1.4 Tesla for not-irradiated sensor.
The solid line corresponds to the parabola fit. Model prediction is superimposed 
(dashed line)}
\label{fig:3a}
\end{figure}

\begin{figure}[htb!]
  \begin{center}
    \leavevmode
    \setlength{\unitlength}{1.0mm}
\includegraphics[clip,scale=0.3]{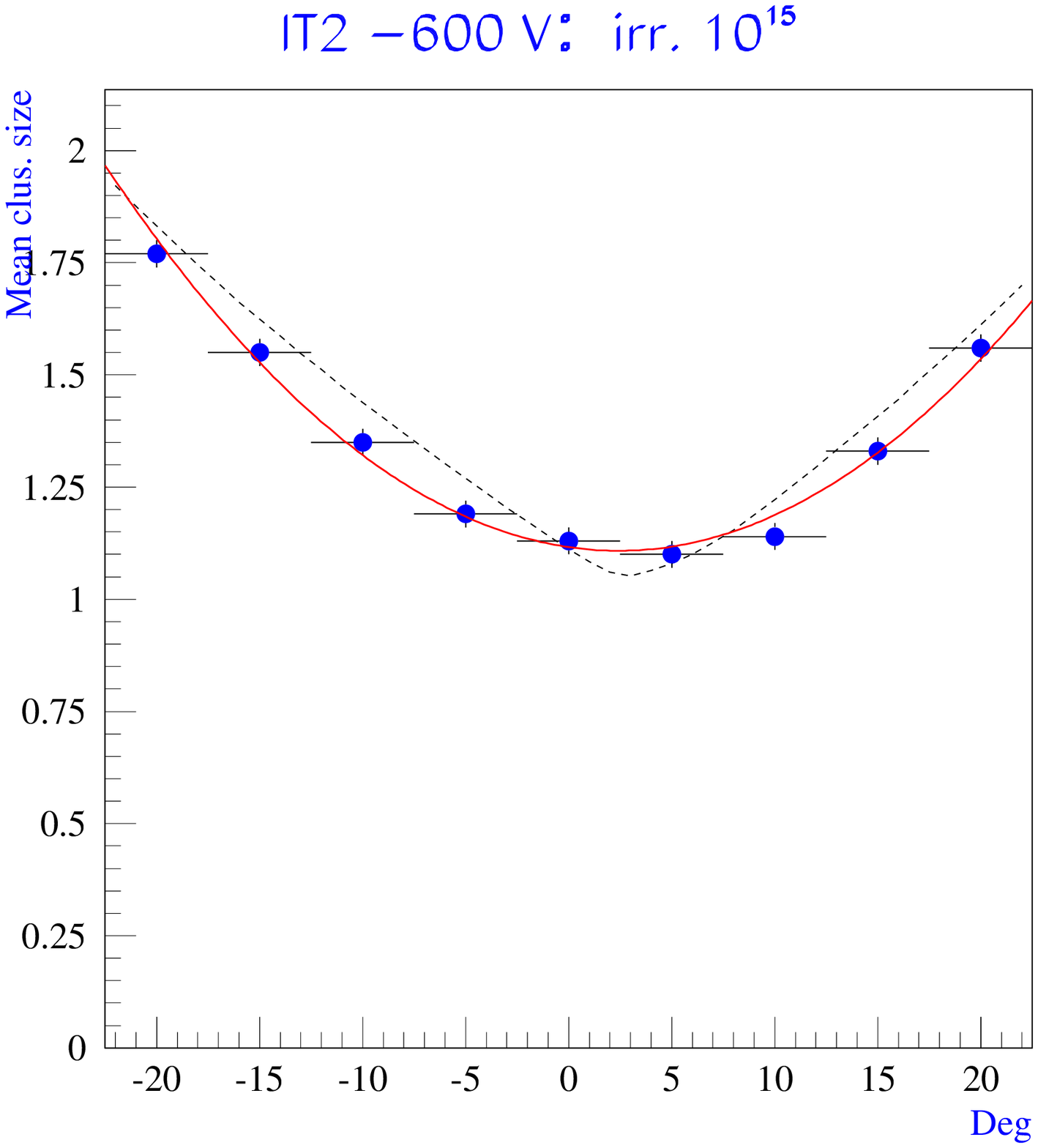}
  \end{center}
\vspace{-0.5cm}
\caption{Mean cluster size as a function of the track incidence angle in 
a magnetic field of 1.4 Tesla for a sensor irradiated to a fluence of $10^{15}$ \nequ/cm\tsup{2}.
The solid line corresponds to the parabola fit. Model prediction is superimposed 
(dashed line)} 
\label{fig:3d}
\end{figure}

The mean cluster size as a function of the angle for not-irradiated sensor operated at 
150 V is shown in fig.~\ref{fig:3a}. The Lorentz angle is 
$9.0^0 \pm 0.4^0 \pm 0.5^0$.
In fig.~\ref{fig:3d} the distribution of the same quantity for a full-irradiated 
sensor is shown. The plot refers to 
data taken in a test beam in 1998. The corresponding Lorentz angle is 
$3.1^0 \pm 0.4^0 \pm 0.6^0$.
The same sensor was measured again in 1999, 
finding a compatible value of $2.7^0 \pm 0.4^0 \pm 0.4^0$. 
An half-irradiated  sensor was also tested at two different operating 
voltages: at 600 V as foreseen during data taking in ATLAS and at the lower
voltage of 150 V for a better understanding of its behaviour.
The measured Lorentz angle values were respectively 
$2.6^0 \pm 0.2^0 \pm 0.3^0$ and $5.9^0 \pm 1.0^0 \pm 0.3^0$. 

The expected mean multiplicity as a function of the track 
incidence angle was numerically calculated, using the properties 
and the operating conditions of the detectors under study 
(temperature, magnetic field, bias voltage, depletion, 
geometry and thresholds). 
This calculation is based on a model which describes charge drifting in
silicon devices.
To compute the drifting trajectories is necessary to know the 
electric field, the magnetic field and the mobility.

The Lorentz angle $\gamma$  is given by \cite{Sho50}

\begin{equation}
\tan\gamma=\mu_H B=r\mu_d B
\end{equation}

where $\mu_H$ is the Hall mobility and $\mu_d$ the drift mobility. 
Their ratio r (Hall factor) depends on the scattering mechanism.
It has a weak dependence on temperature while it does not depend on doping 
level as long as the doping level is below $10^{14} \;\; \mbox{cm}^{-3}$.
The mobility depends on temperature and electric field. 
This dependence was 
parametrized as in \cite{Jac77}. 

\begin{figure}[htbp!]
  \begin{center}
    \leavevmode
    \setlength{\unitlength}{1.0mm}
    \begin{picture}(75,50)
    \put(-6,0){\mbox{\epsfxsize=7.5cm\epsffile{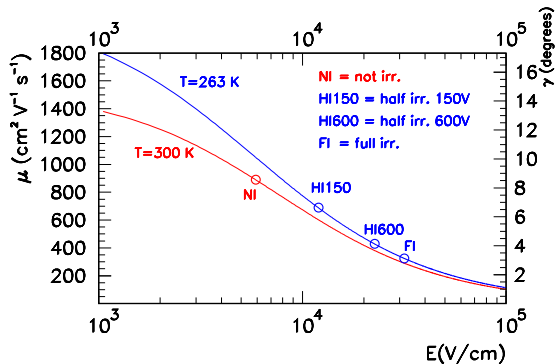}}}  
 \end{picture}
\end{center}
\vspace{-.5cm}
\caption{Drift mobility (left scale) and Lorentz angle (right scale) as a 
function of electric field for 263 K and 300 K.
The points corresponding to the mean electric field and temperature 
of the studied detectors are also reported.}
\label{fig:Mobility}
\end{figure}

Fig. \ref{fig:Mobility} shows the mobility and the Lorentz angle 
(assuming $r=1.2$ and $B=1.4$ T) as a function of the electric field
for $T$=263 K and $T$=300~K (irradiated and not-irradiated detectors). 
The markers correspond to the temperature and the 
{\em mean} electric field $<\mathrm{E}>=\mathrm{V}/\mathrm{d}$ present in the detectors under study,
where V is the applied voltage and d is the depletion depth. 

The irradiated samples have a lower Lorentz angle 
because of lower mobility due to a larger electric field. Actually the electric
field is not constant in the detector due to the presence of spatial 
charges, then Lorentz angle varies throughout the detector with 
the position inside the depletion zone. 
The track incidence angle for which the mean hit multiplicity 
is minimum was defined {\em effective Lorentz angle}. 

The doping concentration was assumed to be uniform. 
This is in agreement with 
the scaling of depletion depth with bias voltage observed in half-irradiated 
detectors: the depletion approximately doubles when the voltage is increased 
from 150~V to 600 V.
With this assumption the electric field varies linearly from the backplane 
to the read-out plane, both in irradiated samples
, which have a p-type substrate and are partially depleted, 
and in not-irradiated devices which are over depleted.

The mean multiplicity for a given detector and track incidence angle  
was computed taking into account charge drifting, charge diffusion
and threshold.  
Experimental results were compared with the model. 
The values of depletion depth used in the model were  
experimentally measured (see Section\ref{sec:dep}).
The results obtained are summarized in table ~\ref{table:2}.

\section{Measurement of the Depletion Depth}
\label{sec:dep}

\begin{figure}[htb]
  \begin{center}
    \leavevmode
    \setlength{\unitlength}{1.0mm}
\includegraphics[clip,scale=0.32]{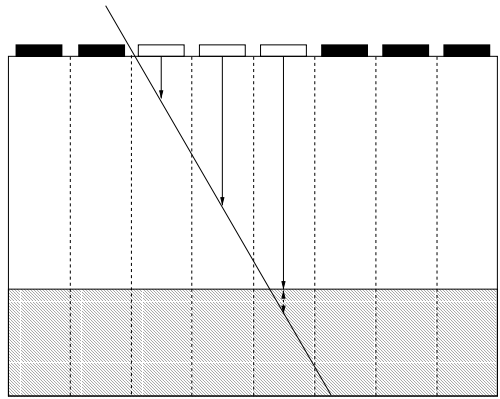}  
  \end{center}
\caption{Schematic view of an irradiated sensor crossed 
by a track. The hatched zone corresponds to the non depleted zone. 
The depth corresponding to each fired pixel is shown.}
\label{fig:pixel3}
\end{figure}

The measurement of the depletion depth was performed 
rotating the detector around the axis parallel to the 
longer size of pixels.   
The angle between the beam direction and the 
normal to the sensor plane was set to $20^0$ or $30^0$.  
The method is illustrated in fig~\ref{fig:pixel3}: a cluster 
of contiguous pixels is activated by the beam particles crossing 
the detector. 

\begin{figure}[htb]
  \begin{center}
    \leavevmode
    \setlength{\unitlength}{1.0mm}
    \begin{picture}(75,75)
      \put(-6,0){\mbox{\epsfxsize=7.5cm\epsffile{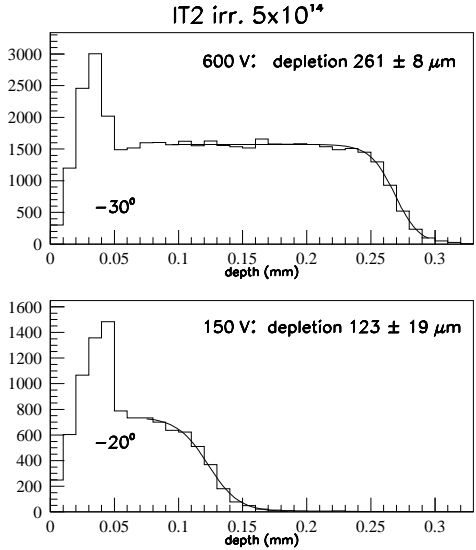}}} 
     \end{picture}
  \end{center}
\vspace{-.5cm}
\caption{ Distribution of the track-segment depth for sensors irradiated to a 
fluence of $5\times10^{14} \mathrm{n_{eq}/cm^2}$ operating at two different 
voltages. The fitting function is superimposed.}
\label{fig:scan}
\end{figure}
The charge collected at each pixel is proportional 
to the length of the track segment in the depleted area below the 
pixel itself.
The alignment of the tracks is performed by the measurement of 
Time Over Threshold of the first fired pixel in the cluster. 
For each fired pixel in the cluster the depth of the center of the segment 
is calculated.

In fig.~\ref{fig:scan} the distributions of the track-segment depth are shown for a sensor irradiated to a fluence of 
$5 \times 10^{14} \mathrm{n_{eq}/cm^2}$ and operated at two 
different values of bias voltage.
The maximum depth is measured determining with a fit the point of inflexion 
of the track-segment depth distribution.

The consistency of the method was checked with not-irradiated 
sensors, that are fully depleted, for which the depletion is 
equal to the nominal thickness.

\section{Conclusions}
The Lorentz angle of ATLAS pixel radiation-hard sensors in a field of
1.4 Tesla was measured.   
For a not-irradiated sensor operated at 150 V a value of  
$9.0^0 \pm 0.4^0 \pm 0.5^0$ was found. 
Two different measurements for a sensor irradiated with 
$10^{15} \mathrm{n_{eq}/cm^2}$ and operated at 600 V gave 
the compatible values of $3.1^0 \pm 0.4^0 \pm 0.6^0$.
and $2.7^0 \pm 0.4^0 \pm 0.4^0$. 
The Lorentz angle depends through the mobility upon the electric field
inside sensors. 
The observed behaviour is well explained by a model based on 
charge drift in silicon.
At the operating conditions for ATLAS pixel sensors
a Lorentz angle of $13^0$ at the beginning of data taking, decreasing to 
$4^0$ after 10 years of operation, is expected.

\end{document}